\input harvmac
\newcount\figno
\figno=0
\def\fig#1#2#3{
\par\begingroup\parindent=0pt\leftskip=1cm\rightskip=1cm\parindent=0pt
\global\advance\figno by 1
\midinsert
\epsfxsize=#3
\centerline{\epsfbox{#2}}
\vskip 12pt
{\bf Fig. \the\figno:} #1\par
\endinsert\endgroup\par
}
\def\figlabel#1{\xdef#1{\the\figno}}
\def\encadremath#1{\vbox{\hrule\hbox{\vrule\kern8pt\vbox{\kern8pt
\hbox{$\displaystyle #1$}\kern8pt}
\kern8pt\vrule}\hrule}}

\overfullrule=0pt

%
\def\underarrow#1{\vbox{\ialign{##\crcr$\hfil\displaystyle
 {#1}\hfil$\crcr\noalign{\kern1pt\nointerlineskip}$\longrightarrow$\crcr}}}
%

\def\bar{\overline}

\def\inbar{\vrule height1.5ex width.4pt depth0pt}
\def\IC{\relax\hbox{\kern.25em$\inbar\kern-.3em{\rm C}$}}
\def\IR{\relax\hbox{\kern.25em$\inbar\kern-.3em{\rm R}$}}
\def\IZ{\relax\ifmmode\hbox{Z\kern-.4em Z}\else{Z\kern-.4em Z}\fi}

\font\zfont = cmss10 

\def\bigone{\hbox{1\kern -.23em {\rm l}}}
\def\ZZ{\hbox{\zfont Z\kern-.4emZ}}


\def\drawbox#1#2{\hrule height#2pt
        \hbox{\vrule width#2pt height#1pt \kern#1pt
              \vrule width#2pt}
              \hrule height#2pt}

\def\Asym#1#2{\vcenter{\vbox{\drawbox{#1}{#2}
              \kern-#2pt       
              \drawbox{#1}{#2}}}}

\batchmode
  \font\bbbfont=msbm10
\errorstopmode
\newif\ifamsf\amsftrue
\ifx\bbbfont\nullfont
  \amsffalse
\fi
\ifamsf
\def\IR{\hbox{\bbbfont R}}
\def\IC{\hbox{\bbbfont C}}

\def\IZ{\hbox{\bbbfont Z}}


\midinsert
\endinsert


\nref\bayen{F. Bayen, M. Flato, C. Fronsdal, A. Lichnerowicz and D.
Sternheimer, {\it Ann. Phys.} {\bf 111}, (1978) 61; {\it Ann. Phys.} {\bf
111} (1978)
111.}

\nref\fedosov{ B. Fedosov, {\it J. Diff. Geom.} {\bf 40} (1994) 1213; {\it
Deformation Quantization and Index Theory} (Akademie Verlag, Berlin
1996).}

\nref\weyl{H. Weyl, {\it Group Theory and Quantum Mechanics}, (Dover, New
York 1931).}

\nref\wigner{E.P. Wigner, {\it Phys. Rev.} {\bf 40} (1932) 749.}

\nref\moyal{J.E. Moyal, {\it Proc. Camb. Phil. Soc.} {\bf 45} (1949) 99.}

\nref\stern{D. Sternheimer, ``Deformation Quantization: Twenty Years
After", in {\it Particles, Fields and Gravitation}, ed.  J.
Rembieli\'nski (American Institute of Physics, New York 1998).}

\nref\zachos{C.K. Zachos, ``Deformation Quantization: Quantum Mechanics
Lives and Works in Phase Space'', hep-th/0110114.}

\nref\berezinone{F.A. Berezin, {\it Izvestiya AN USSR}, ser. math. {\bf 6}
(1972) 1117.}

\nref\berezintwo{F.A. Berezin, {\it Izvestiya AN USSR}, ser. math. {\bf
38} (1974) 1116.}

\nref\berezinthree{F.A. Berezin, {\it Commun. Math. Phys.} {\bf 40} (1975)
153.}

\nref\berezinfour{F.A. Berezin, {\it Izvestiya AN USSR}, ser. math. {\bf
39} (1975) 363.}

\nref\berezinfive{F.A. Berezin, {\it Commun. Math. Phys.} {\bf 63} (1978)
131.}

\nref\schrodinger{F.A. Berezin and M.A. Shubin, {\it The Schr\"odinger
Equation}, (Kluwer Academic Publishers, Dordrecht 1991).}

\nref\rcgone{J. Rawnsley, M. Cahen and S. Gutt, {\it J. Geom. Phys.} {\bf
7} (1990) 45.}

\nref\rcgtwo{ M. Cahen, S. Gutt and J. Rawnsley, {\it Trans. Amer. Math.
Soc.} {\bf 337} (1993) 73; {\it Lett. Math. Phys.} {\bf 30} (1994) 291;
{\it Lett. Math. Phys.} {\bf 34} (1995) 159.}

\nref\kara{A.V. Karabegov, {\it Func. Anal. Appl.} {\bf 30} (1996) 142.}

\nref\martin{M. Schlichenmaier, ``Berezin-Toeplitz Quantization of Compact
K\"ahler Manifolds", in {\it Quantization, Coherent States and Poisson
Structures}, Proceedings of the XIV Workshop on Geometric Methods in
Physics, Bia\l owie$\dot{z}$a, July 9-15 1995, Eds. A. Strasburger et al.
(Polish Scientific Publishers PWN, Warszawa 1998).}

\nref\martinone{M. Schlichenmaier, ``Berezin-Toeplitz Quantization and
Berezin's Symbols for Arbitrary Compact K\"ahler Manifolds'', in {\it
Coherent States, Quantization and Gravity}, Proceedings of the XVII
Workshop on Geometric Methods in Physics, Bia\l owie$\dot{z}$a, July 3-9
1998, Eds. M. Schlichenmaier et al. (Polish Scientific Publishers PWN,
Warszawa 2001).}

\nref\martintwo{M. Schlichenmaier, ``Deformation Quantization of Compact
K\"ahler Manifolds by Berezin-Toeplitz Quantization'', in {\it Quantization, Deformations and Symmetries},
Conf\'erence Mosh\'e Flato 1999, Eds. G. Dito and D. Sternheimer (Kluwer Academic Publishers, Dordrecht 2000).}

\nref\martinthree{A. Karabegov and M. Schlichenmaier, ``Identification of 
Berezin-Toeplitz Deformation Quantization", math.QA/0006063.}

\nref\takhtajan{N. Reshetikhin and L.A. Takhtajan, ``Deformation
Quantization of K\"ahler Manifolds'', math.QA/9907171.}

\nref\sw{N. Seiberg and E. Witten, {\it JHEP} {\bf 09} (1999) 032.}

\nref\k{M. Kontsevich, ``Deformation Quantization of Poisson Manifolds
I'', q-alg/9709040; {\it Lett. Math. Phys.} {\bf 48} (1999) 35.}

\nref\volovich{A. Volovich, ``Discreteness in de Sitter Space and
Quantization of K\"ahler Manifolds'', hep-th/0001176.}

\nref\avolo{M. Spradlin and A. Volovich, ``Noncommutative Solitons on
K\"ahler Manifolds'', hep-th/0106180.}

\nref\isidro{J.M. Isidro, ``Duality and the Equivalence Principle of
Quantum Mechanics'', hep-th/0009221.}

\nref\mielnik{B. Mielnik, {\it Commun. Math. Phys.} {\bf 31} (1974) 221.}

\nref\kibble{T.W.B. Kibble, {\it Commun. Math. Phys.} {\bf 65} (1979)
189.}

\nref\page{D. Page, {\it Phys. Rev. A} {\bf 36} (1987) 3479.}

\nref\anandan{J. Anandan, {\it Found. Phys.} {\bf 21} (1991) 1265.}

\nref\anandantwo{J. Anandan, ``Reality and Geometry of States and
Observables in Quantum Theory'', in {\it International Conference on  
Non-accelerator Particle Physics ICNAPP}: Proceedings, Ed. R. Cowsik
(World Scientific, River Edge N.J. 1994).}

\nref\hugh{L.P. Hughston, ``Geometric Aspects of Quantum Mechanics'', in
{\it Twistor Theory}, Ed. S. Huggett (Marcel Dekker, New York 1995).}

\nref\ashtekar{A. Ashtekar and T.A. Schilling, ``Geometrical Formulation 
of Quantum Mechanics'', in {\it On Einstein's Path}, Essays in Honor of
Engelbert Schucking, Ed. A. Harvey (Springer-Verlag, New York 1999).}

\nref\brody{B.D.C. Brody and L.P. Hughston, {\it J. Geom. Phys.} {\bf 38}
(2001) 19.}

\nref\mi{J.M. Isidro, ``The Geometry of Quantum Mechanics'',
hep-th/0110151.}

\nref\koba{S. Kobayashi, {\it Trans. Amer. Math. Soc.} {\bf 92} (1959)
267.}

\nref\hatfield{B. Hatfield, {\it Quantum Field Theory of Point Particles
and Strings} (Addison-Wesley, New York 1992).}

\nref\dito{J. Dito, {\it Lett. Math. Phys.} {\bf 20} (1990) 125; {\it 
Lett. Math. Phys.} {\bf 27} (1993) 73.}

\nref\cz{ T. Curtright and C. Zachos, {\it J. Phys. A: Math. Gen.} {\bf
32}, (1999) 771.}

\nref\fields{H. Garc\'{\i}a-Compe\'an, J.F. Pleba\'nski, M. Przanowski and
F.J. Turrubiates, {\it Int. J. Mod. Phys. A} {\bf 16} (2001) 2533.}

\nref\nomi{S. Kobayashi and K. Nomizu, {\it Foundations of Differential
Geometry}, (Interscience Publishers, New York 1969) Vol. II.}

\nref\wells{R.O. Wells, {\it Differential Analysis on Complex Manifolds}, 
(Springer-Verlag, Berlin, New York 1980).}

\nref\grad{I.S. Gradshteyn and I.M. Ryzhik, {\it Table of Integrals,
Series and Products} (Academic, New York 1980).}

\nref\connes{A. Connes, M. Flato and D. Sternheimer, {\it Lett. Math.
Phys.} {\bf 24} (1992) 1-12.}

\nref\weinberg{S. Weinberg, {\it Ann. Phys.} {\bf 194} (1989) 336.}

\nref\mielniktwo{B. Mielnik, {\it Phys. Lett. A} {\bf 289} (2001) 1.}

\vskip - 1.5truecm
\Title{hep-th/0112049, CINVESTAV-FIS-01/097}
{\vbox{\centerline{}
\vskip - 2truecm
\centerline{Deformation Quantization of Geometric}
\smallskip
\centerline{Quantum Mechanics}}}
\centerline{H.
Garc\'{\i}a-Compe\'an,$^{a}$\foot{E-mail address: {\tt compean@fis.cinvestav.mx}} J.F. 
Pleba\'nski,$^a$\foot{E-mail address: {\tt pleban@fis.cinvestav.mx}} M.
Przanowski$^{b,a}$\foot{E-mail address: {\tt przan@fis.cinvestav.mx}} and
F.J. Turrubiates$^a$\foot{E-mail address: {\tt fturrub@fis.cinvestav.mx}} }
\smallskip
\centerline{\it $^a$Departamento de F\'{\i}sica}
\centerline{\it  Centro de Investigaci\'on y de Estudios Avanzados del
IPN}
\centerline{\it Apdo. Postal 14-740, 07000, M\'exico D.F., M\'exico}
\smallskip
\centerline{\it $^b$Institute of Physics}
\centerline{\it Technical University of \L \'od\'z}
\centerline{\it W\'olcza\'nska 219, 93-005, \L \'od\'z, Poland}
\vskip 1.5truecm
\noindent
Second quantization of a classical nonrelativistic one-particle system as a
deformation quantization of the Schr\"odinger spinless field is
considered. 
Under the assumption that the phase space of the Schr\"odinger field
is $\IC^{\infty}$, both, the Weyl-Wigner-Moyal and Berezin deformation
quantizations are discussed and compared. Then the geometric quantum
mechanics is also quantized using the Berezin method under the assumption
that the phase space is ${\IC}P^{\infty}$ endowed with the Fubini-Study
K\"ahlerian metric. Finally, the Wigner function for an arbitrary particle
state and its evolution equation are obtained. As is shown this new
``second quantization" leads to essentially different results than the
former one.
For instance, each state is an eigenstate of the total number particle
operator and the corresponding eigenvalue is always ${1 \over \hbar}$.



\Date{December, 2001}

\newsec{Introduction} 

Deformation quantization and, in general, noncommutative geometry have
been matter of a great deal of renewed interest. In deformation
quantization approach the quantization is considered as a noncommutative
deformation ${\cal A}_N$ of the algebra of the classical observables
${\cal A}_C$ in the phase space \refs{\bayen,\fedosov}.
The resulting quantum algebra of linear operators is now equivalent to the
deformation of the original algebra ${\cal A}_C$.  This can be done at the
level of the product of observables ${\cal O}_1 * {\cal O}_2$ or at the
level of the Poisson-Lie bracket $\{{\cal O}_1, {\cal O}_2 \}_*$. It
depends on the type of the deformation quantization which is
involved. In the present paper we consider both types of situations.
Here we deal with the Weyl-Wigner-Moyal deformation quantization
\refs{\weyl - \moyal} (for a recent review of this topic, see
\refs{\stern,\zachos}) and Berezin's one
\refs{\berezinone - \takhtajan}.

For noncommutative geometry the situation is quite similar. In this case
the deformed space is the noncommutative spacetime and the usual algebra
of smooth functions is deformed into an associative and
noncommutative
algebra with the corresponding Moyal $*$-product. Yang-Mills gauge
theories can be transformed into noncommutative gauge theories by
replacing the usual matrix product by the Moyal $*$-product. These
theories are strongly motivated since they can be obtained from the
operator product expansion of string theories \refs{\sw}.

Weyl-Wigner-Moyal deformation quantization is very useful for the
description of {\it flat} finite dimensional phase spaces (or spacetimes)
and many results have been obtained mainly by using this formalism.
However, for the more general phase spaces (or spacetimes) further
generalizations are required. One of them is the Fedosov deformation
quantization \refs{\fedosov} for an arbitrary symplectic manifold of
finite dimension or Kontsevich's deformation quantization \refs{\k} for
the case of a general finite dimensional Poisson manifold. Another
approach, extensively used in our paper, is the Berezin deformation
quantization which is especially useful for K\"ahlerian manifolds.

(It is worth to note the promising application of the Berezin formalism
to the noncommutative sphere and noncommutative solitons on K\"ahler
manifolds \refs{\volovich,\avolo}. Recently, Berezin's deformation
quantization has also been used to construct a nonperturbative formulation
of quantum mechanics which includes $S$-duality symmetries observed in
quantum theories of fields and strings. Such a formulation is based on a
topological limit of the Berezin quantization of the upper
half-plane \refs{\isidro}). 

The aim of the present paper is to apply the Berezin approach to quantize
geometric quantum mechanics and then to compare the result with the usual
second quantization of the Schr\"odinger field.  The geometric
interpretation of quantum mechanics is a subject considered in the
literature by a series of authors \refs{\mielnik - \mi} and is based on
the identification of the quantum phase space coming from the formal
solution of the Schr\"odinger equation for a two state system with the
complex projective space $\IC P^1 \cong {\bf S}^2$. This does admit an
immediate generalization to $\IC P^n$ and in the general case we have to
deal with $\IC P^{\infty}$. All these spaces endowed with the well known
Fubini-Study metrics are, of course, K\"ahler manifolds. (The geometric
structure of $\IC P^{\infty}$ as a K\"ahler manifold has been discussed
for example, by Kobayashi \refs{\koba}). Moreover, the usual axiomatic
formulation of quantum mechanics can be translated into a geometric
language. For instance, the probability transition is given in terms of
the Fubini-Study metric, while the quantum evolution equation is governed
by the K\"ahler form.

Hence, it seems to be natural to consider the geometric quantum mechanics
as a classical theory on the phase space (symplectic manifold
${\IC}P^{\infty}$). The only essential difference between the geometric
quantum mechanics and other classical theories is that in the former one
not every real function on the phase space is an observable since each
observable must be here the expected value of some hermitian operator
(see the formula (3.1)). Consequently, the product of two observables in
general is no longer an observable. However the Poisson bracket of these
observables is still an observable (see Sec. 3).

Now the question is if the quantization of this theory is equivalent to
the usual second quantization. As is shown in the present paper it is not
so, and the quantization of geometric quantum mechanics leads to some new
results which are not observed in the case of the second quantization.

Our paper is organized as follows. In Section 2 we deal with the second
quantization as a deformation quantization of the Schr\"odinger field.
Assuming that the respective phase space is $\IC^{\infty}$ we first use
the Weyl-Wigner-Moyal formalism and then the Berezin one. Section 3 is
devoted to a brief review of the geometric formulation of quantum
mechanics following Refs. \refs{\mielnik - \mi}. We provide here the
notation which will be used in the next sections. Sections 4 and 5 are the
main parts of the paper. In Section 4 the Berezin quantization of the
geometric quantum mechanics is given and some physical results are
obtained which are drastically different from the ones known in the usual
second quantization. In Section 5 we find the Wigner functions for the
particle states. The von Neumann-Liouville evolution equation for an
arbitrary Wigner function is also given. Final remarks (Section 6) close
the paper.

\vskip 2truecm

\newsec{Second Quantization as a Deformation Quantization}

We deal with a nonrelativistic particle without spin in $3$ dimensions
which is moving under the potential $V(\bf{x})$, $\bf{x} \in
\IR^3$. The evolution equation is the usual Schr\"odinger equation

\eqn\uno{i \hbar {\partial \Psi({\bf x},t) \over \partial t} = \bigg(-{ {\hbar}^2 \over 2m} \Delta +
V({\bf x}) \bigg) \Psi({\bf x},t).}

As it is used in the second quantization procedure \refs{\hatfield}, the
Schr\"odinger equation \uno\ is treated as the classical field equation
which can be derived from the following action

$$
S= \int dt L (t),
$$
\eqn\dos{L(t)=\int d^3x \bar{\Psi} \bigg(i \hbar \dot{\Psi} - V \Psi +
{\hbar^2 \over 2m} \Delta \Psi)\bigg),}
where $\dot{\Psi}= {\partial \Psi \over \partial t}$. Then ${\delta S
\over \delta \bar{\Psi}}=0$ is equivalent
to the Schr\"odinger equation \uno\ and ${\delta S \over \delta \Psi}=0$ gives the complex conjugation of
the equation \uno. The canonical momentum is defined by

\eqn\tres{\Pi({\bf x},t)={\delta L(t) \over \delta \dot{\Psi}({\bf x},t)}=i\hbar \bar{\Psi}({\bf x},t).}

For the fundamental Poisson brackets we obtain 

\eqn\cuatro{\{\Psi({\bf x},t),\Pi({\bf x}',t)\}=\delta({\bf x} - {\bf x}')  \Longrightarrow \{\Psi({\bf
x},t),\bar{\Psi}({\bf x}',t) \}= {1 \over {i \hbar}} \delta({\bf x}-{\bf x}'). }

Finally, the Hamiltonian is given by

\eqn\cinco{H={1 \over i \hbar} \int d^3x \Pi \bigg(V \Psi - {\hbar^2 \over 2m} \Delta \Psi\bigg).}

The energy eigenfunctions of the particle are found from the equation

\eqn\seis{\bigg(-{\hbar^2 \over 2m} \Delta + V \bigg) \psi_k({\bf x}) = 
\varepsilon_k \psi_k({\bf x}),}
where $\psi_k({\bf x},t) = \psi_k({\bf x},0) \exp \big\{ -{i \over \hbar}
\varepsilon_kt\big\}$ with normalization $ \int d^3x \ \bar{\psi}_k \psi_{k'}
=
\delta_{kk'}$. We can now expand $\Psi ({\bf x},t)$ in terms of those 
eigenfunctions

$$
\Psi({\bf x},t) = \sum_k {1 \over \sqrt{\hbar}} Z_k(t) \psi_k({\bf x}) =
\sum_k {1 \over
\sqrt{\hbar}} Z_k \exp\big\{-i \omega_k t\big\}\psi_k({\bf x}),
$$

\eqn\siete{ 
Z_k = \sqrt{\hbar} \int \bar{\psi}_k({\bf x}) \Psi({\bf x}) d^3x,
}
where $\omega_k = \varepsilon_k/\hbar, \ \bar{\psi}_k({\bf x}) :=
\bar{\psi}_k({\bf x},0)$ and $\Psi({\bf x})
:= \Psi({\bf x},0)$. The Poisson brackets for the $Z$-variables 
can be found from \cuatro\ and \siete\ to be

$$
\{Z_k, Z_{k'} \} =0=\{ \bar{Z}_k, \bar{Z}_{k'} \}, 
$$
\eqn\ocho{
\{ Z_k, \bar{Z}_{k'} \} = -i \delta_{kk'}.}

Define oscillator variables

\eqn\nueve{
Q_k = {1 \over \sqrt{2}} \big( Z_k + \bar{Z}_{k} \big),  \ \ \ \ \ \ \ 
P_k = {i \over \sqrt{2}} \big( \bar{Z}_k - Z_k\big)}
which by \ocho\ satisfy the algebra

$$
\{Q_k, Q_{k'} \} =0=\{P_k,P_{k'} \},
$$
\eqn\diez{
\{Q_k, P_{k'}\} = \delta_{kk'}.
}

The $Z$-variables can be written in terms of the oscillator variables 
as follows

\eqn\once{ Z_k = {1 \over \sqrt{2}} \big(Q_k + i P_k \big), \ \ \ \ \ \ 
\bar{Z}_k = {1 \over \sqrt{2}} \big(Q_k -i P_k \big).}

Hence in terms of the oscillator variables the field function 
$\Psi$ and its conjugate momentum $\Pi$ are given by

$$
\Psi({\bf x}) = {1 \over \sqrt{2 \hbar}} \sum_k \big(Q_k + i P_k \big)
\psi_k({\bf x}),
$$

\eqn\doce{
\Pi({\bf x}) = \sqrt{\hbar \over 2} \sum_k \big(i Q_k + P_k \big)
\bar{\psi}_k({\bf x}).
}

The time evolution of the system in terms of the oscillator variables 
is described by

$$
Q_k(t) = {1 \over \sqrt{2}} \big( Z_k(t) + \bar{Z}_k(t) \big) 
= Q_k \cos (\omega_kt) +  P_k \sin (\omega_k t),
$$

\eqn\trece{
P_k(t) ={i \over \sqrt{2}} \big( \bar{Z}_k(t) - Z_k(t) \big)
= P_k \cos (\omega_kt) - Q_k \sin (\omega_k t).
}
From \doce\ one quickly finds that

$$
Q_k = \sqrt{\hbar \over 2} \bigg( \int d^3x \Psi({\bf x})
\bar{\psi}_k({\bf x}) + {1 \over i \hbar}
\int d^3x \Pi({\bf x}) \psi_k({\bf x}) \bigg), 
$$

\eqn\catorce{ 
P_k = {1 \over \sqrt{2 \hbar}} \bigg( \int d^3x \Pi({\bf x}) \psi_k({\bf
x}) - i \hbar
\int d^3x \Psi({\bf x}) \bar{\psi}_k({\bf x}) \bigg).
}

It seems to be natural to define the phase space of the system considered
by $ {\cal Z}= \{
(Q_0,Q_1,...,P_0,P_1,...):(Q_0,Q_1,...,P_0,P_1,...) \in \IR^\infty \times
\IR^\infty \} $
endowed with the symplectic form 

\eqn\quince{\omega = \sum_k dP_k \wedge dQ_k.}

In terms of complex coordinates $Z_k$ and $\bar{Z}_k$ the symplectic form
$\omega$ reads

\eqn\dieciseis{\omega= -i \sum_k dZ_k \wedge d\bar{Z}_k}
and one can easily recognize it as the K\"ahler form for $\IC^\infty$. The 
K\"ahler potential ${\cal K}$ for this case is given by

\eqn\diecisiete{ {\cal K}= \sum_k Z_k \bar{Z}_k. } 
Writing 

$$
dQ_k = {\sqrt{\hbar \over 2}} \bigg( \int d^3x \bar{\psi}_k({\bf x})
\delta
\Psi({\bf x}) + {1 \over i \hbar} \int d^3x \psi_k({\bf x}) \delta
\Pi({\bf x}) \bigg),
$$
\eqn\dieciocho{ dP_k = {1 \over \sqrt{2 \hbar}} \bigg( \int d^3x
\psi_k({\bf x}) \delta \Pi({\bf x}) - i \hbar
\int d^3x \bar{\psi}_k({\bf x}) \delta \Psi({\bf x}) \bigg), }
we obtain

\eqn\diecinueve{ \omega = \int d^3x \ \delta \Pi({\bf x}) \wedge \delta
\Psi({\bf x})= i \hbar \int d^3 x \ \delta \Psi^*({\bf x})
\wedge \delta \Psi({\bf x}).}

\vskip 1truecm
\subsec{Weyl-Wigner-Moyal Deformation Quantization of the Schr\"odinger
Field}

Now we are prepared to give the deformation quantization of the
Schr\"odinger field. It can be done symilarly as in the case of classical
fields \refs{\dito,\cz,\fields}. First we deal with the
Weyl-Wigner-Moyal deformation quantization.

Let $F_1=F_1(Q,P)$ and $F_2=F_2(Q,P)$ be two functions on the phase space
$\cal{Z}$. The {\it Moyal $*$-product} is defined by 

\eqn\veinte{\big(F_1 *  F_2\big)(Q,P) =  F_1(Q,P) \exp\bigg\{{i\hbar\over
2} \buildrel{\leftrightarrow}\over {\cal P}\bigg\}F_2(Q,P),}
where $\buildrel{\leftrightarrow}\over {\cal P}$ is the Poisson operator

$$
\buildrel{\leftrightarrow}\over {\cal P} := \sum_k
\bigg({{\buildrel{\leftarrow}\over {\partial}}\over
\partial Q_k} {{\buildrel{\rightarrow}\over {\partial}}\over
\partial P_k} - {{\buildrel{\leftarrow}\over {\partial}}\over
\partial P_k} {{\buildrel{\rightarrow}\over {\partial}}\over
\partial Q_k}\bigg)
$$
$$
= i \sum_k \bigg({{\buildrel{\leftarrow}\over {\partial}}\over
\partial \bar{Z}_k} {{\buildrel{\rightarrow}\over {\partial}}\over
\partial Z_k} - {{\buildrel{\leftarrow}\over {\partial}}\over
\partial Z_k} {{\buildrel{\rightarrow}\over {\partial}}\over
\partial \bar{Z}_k}\bigg)
$$
\eqn\veintiuno{
= \int d^3x \ \bigg({{\buildrel{\leftarrow}\over {\delta}}\over
\delta \Psi(x)} {{\buildrel{\rightarrow}\over {\delta}}\over
\delta \Pi(x)} - {{\buildrel{\leftarrow}\over {\delta}}\over   
\delta \Pi(x)} {{\buildrel{\rightarrow}\over {\delta}}\over
\delta \Psi(x)}\bigg).
}

Employing \once\ and \doce\ one can write the Hamiltonian \cinco\ in the
following form

\eqn\veintidos{H= {1 \over 2} \sum_k \omega_k \big( Q_k^2 + P_k^2 \big)=
\sum_k \omega_k \bar{Z}_k Z_k.  }
Then the Heisenberg equation reads 
\eqn\veintitres{\dot{F} = \{F,H \}_M, }
where $\{ \cdot,\cdot \}_M$ stands for the {\it Moyal bracket}
\eqn\veinticuatro{ \{F,G\}_M:= {1 \over i\hbar} (F*G - G*F). }

It is an easy matter to define the Wigner function for any state and it 
can be done in analogous way as for other classical fields (compare with
\refs{\cz,\fields}). For example the Wigner function of the ground
state is defined by
\eqn\veinticinco{Z_k * \rho_0 = 0}
for all $k$.
With the use of \veinte\ and \veintiuno, Eq. \veinticinco\ can be written
as well
\eqn\veintiseis{ Z_k \rho_0 + {\hbar \over 2} {\partial \rho_0 \over
\partial \bar{Z}_k} = 0.}
This equation has the solution

$$
\rho_0 \sim \exp \bigg( -{2 \over \hbar} \sum_k {Z}_k \bar{Z}_k
 \bigg) = \exp \bigg( - {1 \over \hbar} \sum_k \big( Q_k^2 + P^2_k \big)
\bigg) 
$$ 
\eqn\veintisiete{ = \exp \bigg( {2i \over \hbar} \int d^3 x \
\Psi({\bf x}) \Pi({\bf x}) \bigg). } 
Having given the Wigner function for the ground state one can easily
construct Wigner functions for higher states. For example, the Wigner
function for two particles, one of which is in the state $k_1$ and the
second one in the state $k_2$, is given by

\eqn\veintiocho{\rho_{k_1k_2} \sim \bar{Z}_{k_1} \bar{Z}_{k_2} * \rho_0 *
Z_{k_2} Z_{k_1}.}
It is well known that the Weyl-Wigner-Moyal deformation quantization
arises from the Weyl correspondence.
According to this correspondence if $\widehat{F}$ is any operator acting in
the Hilbert space of states then the Weyl symbol $F_W$ of $\widehat{F}$ is
defined by 

\eqn\veintiochob{F_W(Q,P)=Tr\big\{\widehat{\Omega}(Q,P)\widehat{F}\big\},}
where $\widehat{\Omega}(Q,P)$ is the Stratonovich-Weyl quantizer which can be
written in the following form

\eqn\veintiochot{
\widehat{\Omega}(Q,P)=\int \prod_{m}d\xi_m\exp \bigg\{-{i \over
\hbar}\sum_{k}\xi_k P_k \bigg\} |Q-{\xi \over 2}\rangle  \langle Q+{\xi \over
2}|}
or in terms of $\Psi$ and $\Pi$ as the following operator valued
functional

\eqn\veintiochoc{\widehat{\Omega}[\Psi,\Pi]=\int {\cal D}\xi \exp \bigg\{ -{i \over
\hbar}\int d^3x \xi({\bf x})\Pi({\bf x}) \bigg\} |\Psi - {\xi \over
2} \rangle  \langle \Psi+{\xi \over 2}|.}
It is also known that if $F_{W}$ and $G_{W}$ are the Weyl symbols of the
operators $\widehat{F}$ and $\widehat{G}$, respectively, then the Weyl symbol of
the product $\widehat{F}\widehat{G}$ is given by $F_{W}*G_{W}$. (For details see
for example \refs{\fields}).
 
Now we are going to consider the Berezin deformation quantization of 
the Schr\"odinger field.

\vskip 1truecm
\subsec{Berezin Deformation Quantization of the Schr\"odinger Field}

Consider the complex manifold $\IC^{n+1}$ endowed with a K\"ahler metric

\eqn\veintinueve{ds^2=\sum_{k,l=0}^{n} g_{k\bar{l}} (dZ^k \otimes
d\bar{Z}^l +
d\bar{Z}^l \otimes dZ^k),}
which in terms of the K\"ahler potential ${\cal K}={\cal K}(Z,\bar{Z})$
reads

\eqn\treinta{g_{k \bar{l}}= {\partial^2 {\cal K} \over \partial Z^k 
\partial \bar{Z}^l}.}
The corresponding symplectic form is given by

\eqn\treintaiuno{\omega = -i\sum_{k,l=0}^n g_{k \bar{l}} dZ^k \wedge
d\bar{Z}^l}
and it induces a Poisson bracket on the functions of
$C^{\infty}(\IC^{n+1})$

\eqn\treintaidos{\{f,g\}=\sum_{k,l=0}^n \omega^{\bar{l}k} \bigg( {\partial
f \over
\partial \bar{Z}^l} {\partial g \over \partial
Z^k} - {\partial f \over \partial Z^k} {\partial g \over \partial
\bar{Z}^l} \bigg),}
where $\omega^{\bar{l}k}$ is the tensor inverse to the symplectic form,
i.e. $\sum_{l=0}^n \omega^{j\bar{l}} \omega_{\bar{l}k}=\delta^j_k$.
We describe now the Berezin quantization of the classical system on
$\IC^{n+1}$ endowed with a K\"ahler metric
\refs{\berezintwo,\berezinfour,\takhtajan}.

Let $d\mu$ be the volume form on $\IC^{n+1}$

\eqn\treintaitres{d\mu(Z,\bar{Z})= \bigg( {\omega \over 2 \pi \hbar}
\bigg)^n = det(g_{i\bar{j}}) \prod_{k=0}^{n} {dZ^{k} \wedge d\bar{Z}^k 
\over 2\pi i\hbar}.}
Denote now by ${\cal{F}}_{\hbar}$ the Hilbert space of entire
functions on $\IC^{n+1}$, square summable with respect to
the Gaussian measure $\exp\{-{1 \over \hbar}{\cal K}(Z,\bar{Z})\}d\mu(Z,\bar{Z})$.

The inner product of two functions
$f_1,f_2 \in {\cal{F}}_{\hbar}$ is defined by

\eqn\treintaicuatro{(f_1,f_2)=c(\hbar)\int_{\IC^{n+1}}f_1(Z)\bar{f_2(Z)}\exp
\bigg\{-{1 \over \hbar}{\cal K}(Z,\bar{Z}) \bigg\} d\mu(Z,\bar{Z}).} 
Let $\{f_k\}, \  k=1,...$ defines an arbitrary orhonormal basis in
${\cal F}_{\hbar}$ and let 

\eqn\treintaicinco{{\cal
B}(Z,\bar{V})=\sum_{k=1}f_k(Z)\bar{f_k(V)}}
be the ${\it Bergman\  kernel}$. (From the physical point of view the
Bergman kernel will correspond to the coherent states).
Then the holomorphic functions
$\Phi_{\bar V}(Z):={\cal B}(Z,\bar{V})$ parametrized by $\bar{V} \in
\bar{\IC}^{n+1}$, form a supercomplete system in ${\cal F}_{\hbar}$. (Note that the overbar means complex
conjugation and not the closure of the set). For any bounded operator $\widehat{F}$ in ${\cal F}_{\hbar}$ one
defines the following function

\eqn\treintaisesis{F_B(Z,{\bar V})= {(\widehat{F}\Phi_{\bar V},\Phi_{\bar
Z})
\over (\Phi_{\bar V},\Phi_{\bar Z})}.}
The function $F_B(Z,{\bar Z}) \in C^{\infty}(\IC^{n+1})$ is called the
{\it covariant symbol} of the operator $\widehat{F}$.
Now if $F_B(Z,{\bar Z})$ and $G_B(Z,{\bar Z})$ are two covariant symbols
of $\widehat{F}$ and $\widehat{G}$, respectively, then the covariant symbol of
$\widehat{F}\widehat{G}$ is given by the {\it Berezin-Wick star product} 
$F_B *_B G_B$

$$
(F_B*_BG_B)(Z,{\bar Z})=
$$
$$
c(\hbar) \int_{\IC^{n+1}} F_B(Z,{\bar V})G_B(V,{\bar
Z}) {{\cal B}(Z,{\bar V}){\cal B}(V,{\bar Z}) \over {\cal B}(Z,{\bar Z})}
\exp \bigg\{-{1 \over \hbar} {\cal K}(V,{\bar V}) \bigg\} d\mu(V,{\bar V})
$$
\eqn\treintaisiete{
c(\hbar) \int_{\IC^{n+1}}F_B(Z,\bar{V})G_B(V,\bar{Z}) \exp \bigg\{ {1
\over \hbar} {\cal K}(Z,\bar{Z};V,\bar{V}) \bigg \} d\mu(V,\bar{V}) }
where ${\cal K}(Z,\bar{Z};V,\bar{V}):= {\cal K}(Z,\bar{V})+{\cal
K}(V,\bar{Z})-{\cal K}(Z,\bar{Z})-{\cal K}(V,\bar{V})$ is called the {\it 
Calabi diastatic function}.
One can also show that

\eqn\treintaiocho{Tr \ \widehat{F}=c(\hbar) \int_{\IC^{n+1}} F_B(Z,{\bar Z})
{\cal B}(Z,{\bar Z}) \exp \bigg\{-{1 \over \hbar} {\cal K}(Z,{\bar Z}) \bigg\}
d\mu(Z,{\bar
Z}). }
In order to specialize the Berezin deformation quantization to the case of
the Schr\"odinger field we should assume that our complex space is
infinite dimensional i.e. we deal with $\IC^{\infty}$ and the metric
is given by $g_{k{\bar l}}=\delta_{kl}$. The K\"ahler function is
therefore defined by (2.17).
In this case $c(\hbar)=1$ and the orthonormal basis in the Hilbert space
${\cal F}_{\hbar}$ can be chosen to be the Fock basis

\eqn\treintainueve{f_{(s_0,s_1,...)}(Z)=\prod_{l} {Z_l^{s_l} \over
\sqrt{s_l! \hbar^{s_l}}}.} 
For the Bergman kernel \treintaicinco\ we obtain now 

\eqn\cuarenta{{\cal B}(Z,{\bar V}) = \exp \bigg\{ {1 \over \hbar} \sum_k 
Z_k {\bar V}_k \bigg\} = \exp \bigg\{ {1 \over \hbar} {\cal K}(Z,{\bar V})
\bigg\}.
}
Straightforward calculations show that the covariant symbol of the
operator 

\eqn\cuarentaiuno{\widehat{F}=\sum_{j,k,l,m} F_{jk}^{lm}(\widehat
{a}^{\dag}_l)^j(\widehat{a}_m)^k,}
where $\widehat{a}^{\dag}_l$ and $\widehat{a}_m$ are the creation and
annihilation operators, respectively, reads

\eqn\cuarentaidos{F_B(Z,{\bar Z})= 
\sum_{j,k,l,m}F_{jk}^{lm}(Z_l)^j(\bar{Z}_m)^k = F_{Wick}(\bar{Z},Z).}
Here $F_{Wick}(Z,\bar{Z})$ stands for the {\it Wick symbol} of
the operator (2.44) \refs{\berezinone,\berezintwo,\schrodinger} (Note the
order of the arguments $Z,\bar{Z}$).

As can be proved (see e.g. \refs{\berezintwo}) the relation between
covariant
$F_B$ and the Weyl $F_W$ symbols of operator $\widehat{F}$ reads

$$
F_W={\cal N} F_B(\bar{Z},Z)={\cal N}F_{Wick}(Z,\bar{Z})
$$
$$
{\cal N} = \exp \bigg\{- {\hbar \over 2} \sum_k {\partial^2 \over
\partial Z_k \partial {\bar Z}_k} \bigg\}
$$

$$
=\exp \bigg\{- {\hbar \over 4} \sum_k ({\partial^2 \over \partial
Q_k^2} + {\partial^2 \over \partial P_k^2}) \bigg\}
$$
\eqn\cuarentaitres{
=\exp \bigg\{- {i \hbar \over 2} \int d^3x {\delta^2 \over \delta \Psi({\bf
x}) \delta \Pi({\bf x})} \bigg\}.
}
Consequently, the Moyal and Berezin-Wick star products are related by

$$
F*G= {\cal N} \bigg({\cal N}^{-1}F *'_B {\cal N}^{-1} G\bigg)
$$
\eqn\cuarentaicuatro{
F *'_B G= {\cal N}^{-1}\bigg({\cal N} F * {\cal N} G\bigg),
}
where
$$
F(Z,\bar{Z})*'_BG(Z,\bar{Z}):=c(\hbar)
\int_{\IC^{n+1}}F(V,\bar{Z})G(Z,\bar{V}) \exp \bigg\{ {1 \over \hbar}
{\cal K}(Z,\bar{Z};V,\bar{V}) \bigg\} d\mu(V,\bar{V}) 
$$
\eqn\cuarentaicuatrob{ =G(Z,\bar{Z})*_BF(Z,\bar{Z}).}
In what follows $*'_B$-product will be also called the {\it Berezin-Wick
star product} $*'_B$. 
\vskip 2truecm
\newsec{Geometric Quantum Mechanics}

As have been pointed out by many authors \refs{\mielnik - \mi},
quantum mechanics can be formulated as a geometric theory on a symplectic
manifold. We would like to explain briefly this approach.

States of a quantum mechanic system are represented by rays in the
associated infinite dimensional Hilbert space ${\cal H}$. The expectation
value of an observable $\widehat{\widehat{F}}$ in a state defined by the
ket vector $|Z \rangle =|Z_0,Z_1,...\rangle $ is given by 

\eqn\cuarentaicinco{ \langle \widehat{\widehat{F}}\rangle = { \langle
Z|\widehat{\widehat{F}}|Z \rangle \over \langle Z|Z\rangle }.}
Henceforth we use double hat for operators in usual quantum mechanics and
single hat for operators acting in the Hilbert space of field states. 
The expression (3.1) suggests that the space of rays in ${\cal H}$
i.e. the complex projective space ${\IC}P^{\infty}$ represents the phase
space of the system and the observables are the functions on ${\IC}P^{\infty}$
of the form \cuarentaicinco. The complex coordinates $Z_k$ introduced in
the previous section (see \siete) constitute the {\it homogeneous
coordinates} of ${\IC}P^{\infty}$. Let $\widetilde{U}_j$ be a subset of
${\IC}^{\infty}$ defined by $U_j=\{(Z_0,Z_1,...) \in {\IC}^{\infty} : Z_j
\not= 0 \}$.
Then one can define the {\it inhomogeneous coordinates} on the respective
coordinate neighborhood $U_j \subset {\IC}P^{\infty}$, where $U_j$ is the
projection of $\widetilde{U}_j$ on ${\IC}P^{\infty}$, as follows 
$$
z^0_{(j)} = {Z_0 \over Z_j}, \ \ z^1_{(j)} = {Z_1 \over Z_j}, \ \ ... \ .
$$

In terms of the coordinates $Z$ or $z$ the observable $\langle
\widehat{\widehat{F}} \rangle$ reads

$$
\langle \widehat{\widehat{F}} \rangle = {\sum_{k,l}F_{kl} \bar{Z}_k
Z_l \over \sum_k Z_k \bar{Z}_k} 
$$
\eqn\cuarentaiseis{ 
= { {\sum_{k,l \not=j} F_{kl} \bar{z}^k_{(j)} z^l_{(j)} + \sum_{k \not=j}
(F_{jk} z^k_{(j)}+ F_{kj} \bar {z}^k_{(j)}) + F_{jj} } \over {1 + \sum_{k
\not=j} z^k_{(j)} {\bar z}^k_{(j)} }},} 
where $F_{kl}:= \langle \psi_k|\widehat{\widehat{F}}|\psi_l \rangle =
\bar{F}_{lk}$ for all $l,k.$

In particular from \veintidos\ and \cuarentaiseis\ we get for the
Hamiltonian $\langle \widehat{\widehat{H}} \rangle$

\eqn\cuarentaisiete{\langle \widehat{\widehat{H}} \rangle =
{\sum_{k \not= j}\omega_k
\bar{z}^k_{(j)}z^k_{(j)} + \omega_j \over (1+|z_{(j)}|^2)}, }
where $|z_{(j)}|^2:= \sum_{k \not= j} |z^k_{(j)}|^2$.

The quantum phase space ${\IC}P^{\infty}$ can be endowed in a natural
manner with a Riemannian metric. To this end consider two ket vectors $|Z \rangle$
and $|Z+dZ \rangle$. The transition probability $p(|Z\rangle,|Z+dZ\rangle)$  between
$|Z\rangle$ and $|Z+dZ\rangle$ is given by

\eqn\cuarentaiocho{ p(|Z\rangle,|Z+dZ\rangle) = 
{ \langle Z+dZ|Z\rangle \langle Z|Z+dZ\rangle \over \langle Z|Z\rangle \langle Z+dZ|Z+dZ\rangle}. }
Simple calculations show that up to the second order in $dZ$ the
transition probability $p(|Z\rangle,|Z+dZ\rangle)$ reads

$$
p(|Z\rangle,|Z+dZ\rangle)=1 - { \langle dZ|dZ\rangle \langle Z|Z\rangle- \langle
Z|dZ \rangle \langle dZ|Z\rangle \over
|\langle Z|Z\rangle|^2} 
$$
\eqn\cuarentainueve{
= 1 - \sum_{k,l} {(\sum_m |Z_m|^2) \delta_{kl} - {\bar Z}_k Z_l  \over
\sum_m |Z_m|^2 } dZ_k d{\bar Z}_l.}

The second term of the right hand side of \cuarentainueve\ can be written
in terms of the inhomogeneous coordinates $z^k_{(j)}$ as follows

\eqn\cincuenta{ \sum_{k,l} {(\sum_m |Z_m|^2) \delta_{kl} - {\bar Z}_k Z_l
\over \sum_m |Z_m|^2 } dZ_k d{\bar Z}_l  =
\sum_{k,l\not=j}{(1+|z_{(j)}|^2)\delta_{kl} - {\bar z}^k_{(j)}z^l_{(j)} 
\over (1+|z_{(j)}|^2)^2} dz^k_{(j)}d{\bar z}^l_{(j)}. } 
This suggests us to define the metric $ds^2$ on the quantum phase space
${\IC}P^{\infty}$ such that on any $U_j \in {\IC}P^{\infty}$ $ds^2$ is
proportional to (3.6). For further correspondence between the usual
second quantization and the deformation quantization of geometric quantum
mechanics we take the metric $ds^2$ to be of the form 

$$
ds^2= \sum_{k,l\not=j} g_{k{\bar l}}\bigg(dz^k_{(j)} \otimes d{\bar z}^l_{(j)}
+ d{\bar z}^l_{(j)} \otimes dz^k_{(j)} \bigg),
$$

\eqn\cincuentaiuno{
g_{k{\bar l}}= {(1+|z_{(j)}|^2)\delta_{kl} - {\bar
z}^k_{(j)}z^l_{(j)}) \over (1+|z_{(j)}|^2)^2}, \ \ \ \  k,l \not= j.}

The above metric is up to a constant factor the well known Fubini-Study
metric \refs{\nomi,\wells} and ${\IC}P^{\infty}$ endowed with this
metric is a K\"ahler
manifold. Then the $ds^2$ can be defined on $U_j$ in terms of the K\"ahler
potential ${\cal K}$ as follows
 
$$
g_{k{\bar l}}= {\partial^2 {\cal
K}(z_{(j)},{\bar z}_{(j)}) \over \partial z^k_{(j)} \partial {\bar
z}^l_{(j)}},
$$
\eqn\cincuentaiunob{
{\cal K}= {\cal K}(z_{(j)},{\bar z}_{(j)}) =
\ln{(1 + |z_{(j)}|^2)}= \ln \bigg( \sum_k z_{(j)}^k
\bar{z}_{(j)}^k \bigg).}

It is easy to show that for any $p\in U_j\cap U_l$ of coordinates
$z_{(j)}$ in $U_j$ and $z_{(l)}$ in $U_l$ the following transformation
rule
\eqn\cincuentaiunoc{{\cal K}(z_{(j)},\bar{z}_{(j)})={\cal
K}(z_{(l)},\bar{z}_{(l)})+ 2 \ln |z^l_{(j)}|}
holds.

The K\"ahler form $\Omega$ is defined by 

\eqn\cincuentaidos{\Omega = -i \sum_{k,l\not=j}g_{k{\bar l}} dz^k_{(j)} \wedge
d{\bar z}^l_{(j)} }
for any $j$.
Now we are going to define the symplectic form $\omega = \sum_{k,l\not=j} \omega_{k{\bar
l}} dz^k_{(j)} \wedge d{\bar z}^l_{(j)}$ on the quantum
phase space in such a way that for any function $f$ the evolution 
equation reads

\eqn\cincuentaitres{ {\dot f} =\{f,\langle \widehat{\widehat{H}} \rangle\}
=\sum_{k,l \not=j} \omega^{{\bar k}l}
\bigg( {\partial f \over \partial {\bar z}^k_{(j)}} {\partial \langle 
\widehat{\widehat{H}} \rangle \over \partial z^l_{(j)}} -  {\partial
\langle \widehat{\widehat{H}} \rangle 
\over \partial {\bar z}^k_{(j)}} {\partial f \over \partial z^l_{(j)}}
\bigg). } 
In particular for $z^k_{(j)}$ and ${\bar z}^k_{(j)}$ we have
\eqn\cincuentaicuatro{{\dot z}^k_{(j)}=-\sum_{l \not= j}\omega^{{\bar l}k} 
{\partial \langle \widehat{\widehat{H}} \rangle  \over \partial {\bar
z}^l_{(j)}}, \ \ \ \ \
{\dot{\bar z}}^k_{(j)}=\sum_{l \not= j}\omega^{{\bar k}l} {\partial
\langle \widehat{\widehat{H}} \rangle  \over \partial z^l_{(j)}}, \ \ \ \
k \not= j,
} 
where $\langle \widehat{\widehat{H}} \rangle$ is given by \cuarentaisiete.
However, by
direct calculations one obtains
\eqn\cincuentaicinco{{\dot z}^k_{(j)}={d \over dt} \bigg({Z_k \over
Z_j}\bigg)= iz^k_{(j)}(\omega_j - \omega_k), \ \ \ \ \  {\dot
{\bar z}}^k_{(j)}={d \over dt} \bigg( {\bar {Z_k \over Z_j}} \bigg)=
-i{\bar z}^k_{(j)}(\omega_j - \omega_k). }
(There is no summation over $k$!).

Comparing both expressions \cincuentaicuatro\ and \cincuentaicinco\ we
conclude that 
\eqn\cincuentaiseis{\omega_{k{\bar l}}=-ig_{k{\bar l}}.}
Hence the symplectic form $\omega$ on the quantum phase space compatible
with the evolution equation \cincuentaitres\ is equal to the K\"ahler form
$\Omega$
\eqn\cincuentaisiete{\omega = \Omega.} 
This brief outline of the geometric quantum mechanics shows that from this
point of view quantum mechanics can in a sense be treated as a classical
theory on the infinite dimensional phase space ${\IC}P^{\infty}$.
Therefore, it seems natural to look for quantization of this classical
theory. One expects that such a quantization should be equivalent to the
usual second quantization. But as we are going to demonstrate in the next
section this is not so. This prove will be done with the use of
Berezin's deformation quantization on ${\IC}P^n$ with $n \rightarrow
\infty$.

Here an important comment is needed. The analogy between the geometric
quantum mechanics and classical theory should be considered on the
level of the Poisson-Lie algebra and not on the level of the usual product
algebra of observables. This follows from the fact that the usual product
of two observables $\langle \widehat{\widehat{F}} \rangle \langle
\widehat{\widehat{G}} \rangle$ in general is no longer an observable in a
sense that it cannot be represented in the form of (3.1). From the other
hand, using the formula
\eqn\cincuentaisieteb{\omega^{\bar{k}l}=ig^{\bar{k}l}=i(1+|z|^2)(\delta^{kl}+
\bar{z}^kz^l)} after straightforward calculations one can show that
(compare with \refs{\anandantwo})

\eqn\cincuentaisietec{ \{ \langle
\widehat{\widehat{F}} \rangle, \langle\widehat{\widehat{G}}\rangle \} = -i
\langle [\widehat{\widehat{F}},\widehat{\widehat{G}}] \rangle,}
what means that the Poisson bracket of two observables is also an
observable. Hence, deformation quantization of the geometric quantum
mechanics is rather a deformation of the Poisson-Lie algebra than a
deformation of the usual product algebra. This is so at least in the case
of linear quantum mechanics. The non-linear case will be consider in a
separate paper.     

\vskip 2truecm
\newsec{Berezin's Quantization of Geometric Quantum Mechanics}

We deal with ${\IC}P^n$ endowed with the metric \cincuentaiuno\ defined by
the K\"ahler potential \cincuentaiunob. Then the K\"ahler form
$\Omega$ and the symplectic form $\omega$ are given by \cincuentaidos\
and \cincuentaisiete , respectively.  First, in analogy to the case of the
Berezin quantization on ${\IC}^n$ considered in the section 2.2, we
would like to define the corresponding Hilbert space ${\cal F}_{\hbar}$.
But the obvious problem arises as the only entire function on ${\IC}P^n$
is, according to the Liouville theorem, the constant function. So the
natural idea is to consider ${\cal F}_{\hbar}$ as the space of sections
$Sec({\cal L})$ of some complex line bundle ${\cal L}$ over ${\IC}P^n$
which admits the local trivialization $U_j \times {\IC}$ for any $j$,
$j=0,1,...,n$ \refs{\rcgone,\rcgtwo}. As the measure of the set ${\IC}P^n
- U_j$ is equal to zero for every $j$ one can look for a scalar product in
${\cal F}_{\hbar}$ which by the analogy to \treintaicuatro\ should be
defined as follows

\eqn\cincuentaiocho{(f_1,f_2)=c(\hbar)\int_{U_j}f_{1(j)}(z_{(j)}) 
\bar{f_{2(j)}(z_{(j)})} \exp \bigg\{-{1 \over \hbar} {\cal 
K}(z_{(j)},\bar{z_{(j)}}) \bigg\}
d\mu(z_{(j)},\bar{z_{(j)}}), \ \ \ \ \forall \  j, }
where $f_1, f_2 \in Sec({\cal L})$, $f_{1(j)}$ and $f_{2(j)}$ are the
local representations of $f_1$ and $f_2$, respectively, on $U_j$ and
$d\mu$ is the measure     

$$
d\mu(z_{(j)},\bar{z_{(j)}})= \bigg( {\omega \over 2
\pi \hbar} \bigg)^n = det(g_{i\bar{l}}) \prod_{k\not=j} {dz^{k}_{(j)}
\wedge d\bar{z}^k_{(j)} \over 2\pi i\hbar}
$$
\eqn\cincuentainueve{
=\exp\{-(n+1)ln(1+|z_{(j)}|^2) \} \prod_{k \not= j} {dz^{k}_{(j)}
\wedge d\bar{z}^k_{(j)} \over 2\pi i\hbar}.
}

Now using the formula \cincuentaiunoc\ we can quickly find that the
definition of the scalar product \cincuentaiocho\  is independent of the
index $j$ if and only if the representations of the sections on $U_j$ and
$U_l$ are related by
\eqn\sesenta{f_{(j)}(z_{(j)})=(z^l_{(j)})^{{1 \over  \hbar}}
f_{(l)}(z_{(l)}) }
on $U_j \cap U_l$.
This rule of transformation makes sense only if ${1 \over \hbar}=N$
{\it where $N$ is some positive integer} [10,11,14,15]. We then assume
that indeed it is
so. Consequently our construction indicates that the line bundle
${\cal L}$ is defined by the transition functions

\eqn\sesentab{h_{jl}:U_j \cap U_l \rightarrow {\IC}, \ \ \ \ \ \ \
h_{jl}=(z^l_{(j)})^{1 \over \hbar}.}
Therefore
\eqn\sesentaiuno{{\cal L}=\otimes^{1 \over \hbar} (U_{1,n+1})^{-1},}
where $U_{1,n+1}$ is the {\it universal complex line bundle over}
${\IC}P^n$ \refs{\wells}.
Then the Hilbert space is defined by
\eqn\sesentaidos{{\cal F}_{\hbar}=Sec({\cal L}).}
(For detail analysis of this construction see \refs{\rcgone,\rcgtwo}.)

As the forthcoming calculations will be performed in the open set $U_0$ we
use for simplicity the natural abbreviations by omitting the lower index
$(0)$. So for example we write $z^k := z^k_{(0)}$, $f := f_{(0)}$,
 etc.
First, let's compute the factor $c(\hbar)$ which appears in the definition
of the scalar product \cincuentaiocho. To this end one assumes that the
norm of the cross section of the bundle $\cal{L}$ which on $U_0$
is represented by the unity function $f(z)=1$ is equal to $1$. So
substituting (3.8) and (4.2) into (4.1) and taking also that
$f_1(z)=f_2(z)=1$ we obtain

\eqn\sesentaitres{1=c(\hbar)\int_{U_0} {1 \over (1+|z|^2)^{{1 \over
\hbar}+n+1}} \prod_{k=1}^n {dz^k \wedge d\bar{z}^k \over 2\pi i \hbar}.}
The integral in \sesentaitres\ can be evaluated (see Ref. \refs{\grad},
the integral 4.638-3) to give

$$
\int_{U_0} {1 \over (1+|z|^2)^{{1 \over \hbar}+n+1}} \prod_{k=1}^n {dz^k
\wedge d\bar{z}^k \over 2\pi i \hbar} = \hbar^{-n} {\Gamma({1 \over
\hbar}+1) \over \Gamma({1 \over \hbar}+n+1)}.
$$    
Introducing this result into \sesentaitres\ one finds 

\eqn\sesentaicuatro{c(\hbar)=\hbar^n {\Gamma({1 \over \hbar}+n+1) \over
\Gamma({1 \over \hbar}+1)}.} 
Remember that ${1 \over \hbar}=N\in {\IZ}_+$.
One can check that for any monomial $f(z)$ on $U_0$ of degree greater than
${1 \over \hbar}$ the integral 
$$
c(\hbar)\int_{U_0} |f(z)|^2 {1 \over (1+|z|^2)^{{1 \over
\hbar}+n+1}} \prod_{k=1}^n {dz^k \wedge d\bar{z}^k \over 2\pi i \hbar}
$$
diverges. It means that for $n< \infty$ the dimension of the Hilbert space
${\cal F}_{\hbar}$ is finite.
To proceed further, especially to find the Bergman kernel, we need an
orthonormal basis of ${\cal F}_{\hbar}$.
One expects that an orthonormal basis of ${\cal F}_{\hbar}$ can be
constituted by the sections of the line bundle ${\cal L}$ such that on $U_0$
they are represented by monomials of degree not greater than ${1 \over
\hbar}$. Therefore, consider the monomials on $U_0$ of the following form
$$
e_{(s_1,...,s_n)}(z)=\alpha_{(s_1,...,s_n)}(z^1)^{s_1}...(z^n)^{s_n}, \ \
\ \ s_1+...+s_n \leq {1 \over \hbar}, \ \ s_1,...,s_n \geq 0,
$$    
where $\alpha_{(s_1,...,s_n)}$ is some positive factor.
By straightforward calculations, employing also the formulas (obtained by 
{\it Mathematica})
$$
\sum_{k=0}^s {s \choose k} \Gamma(k+{1 \over 2})\Gamma(s-k+{1 \over 2}) =
\pi s! 
$$ 
and
$$
\sum_{k=0}^s \sum_{l=0}^r (-1)^m {s \choose k} {2r \choose 2l} \Gamma(s +
l - k + {1 \over 2}) \Gamma (r+k -l + {1\over 2}) = 0
$$
one gets  

\eqn\sesentaicuatro{
(e_{(s_1, \dots , s_n)}, e_{(s'_1, \dots , s'_n)}) = \delta_{s_1 s'_1}
\dots \delta_{s_n s'_n}     \Longleftrightarrow 
\alpha_{(s_1,...,s_n)} = \sqrt{{1 \over \hbar}! \over s_1! \dots s_n!
({1\over \hbar} - \sum_{k \not = 0} s_k )!}.}
Hence the set $\big\{ e_{(s_1, \dots , s_n)}(z) = \sqrt{{1 \over 
\hbar}! \over s_1! \dots s_n! ({1\over  \hbar} - \sum_{k \not=0} s_k )!}
(z^1)^{s_1} \dots (z^n)^{s_n} \big\}_{s_1+...+s_n \leq {1\over  \hbar}}$
represents in $U_0$ an orthonormal basis of ${\cal
F}_{\hbar}:\{e_{(s_1,...,s_n)}\}_{s_1+...+s_n \leq {1 \over \hbar}}.$

Now we are in a position to define the Bergman kernel ${\cal B}$ which in
fact should be a global section of the bundle ${\cal L} \otimes \bar{{\cal
L}}$ over ${\IC}P^n \times \bar{{\IC}P^n}$. By the analogy to (2.38) the
representation of the Bergman kernel ${\cal B} \in Sec({\cal L} \otimes
\bar{{\cal L}})$ on $U_0 \times \bar{U_0}$ is defined by

$$
{\cal B}(z,\bar{v}) = \sum_{s_1 + \dots s_n \leq {1 \over \hbar}}
e_{(s_1, \dots , s_n)}(z)\overline{ e_{(s_1, \dots , s_n)}(v)}
$$
\eqn\sesentaicinco{
= {{1\over \hbar}! \over s_1! \dots s_n!
({1\over \hbar} - \sum_{k \not=0} s_k )!} (z^1)^{s_1} \dots (z^n)^{s_n} 
(\bar{v}^1)^{s_1} \dots (\bar{v}^n)^{s_n}
= (1 + \sum_{k=1}^n  z^k \bar{v}^k)^{1 \over \hbar}.}

One quickly finds that the representation of ${\cal B}$ on any $U_j
\times \bar{U_l}$, $j,l=0,...,n$ reads
\eqn\sesentaicincob{
{\cal B}(z_{(j)},\bar{v}_{(l)})=\big( \sum_{k=0}^n z_{(j)}^k
\bar{v}_{(l)}^k \big)^{1 \over \hbar}.
}
Consequently the holomorphic functions $\Phi_{\bar{v}}(z) = {\cal
B}(z,\bar{v})$ on $U_0$ parametrized by $\bar{v} \in \bar{U_0}$ represent 
a supercomplete system in the Hilbert space ${\cal F}_{\hbar}$  i.e.
the set $\{\Phi_{\bar{v}} \in {\cal F}_{\hbar}\}_{\bar{v} \in \bar{U_0}}$
such that $(f,\Phi_{\bar{v}})=f(v)$, for all $f \in {\cal F}_{\hbar}$.
Note that in analogous way one can find a supercomplete system in
${\cal F}_{\hbar}$ parametrized by the points of any $\bar {U_l}$.
This supercomplete system $\{\Phi_{\bar{v}_{(l)}} \in
{\cal F}_{\hbar}\}_{\bar{v}_{(l)} \in \bar{U_l}}$ is defined in terms of
its representation $\{\Phi_{\bar{v}_{(l)}}(z_{(j)})\}_{\bar{v}_l \in
\bar{U_l}}$ on $U_j$ by

\eqn\sesentaicincob{\Phi_{\bar{v}_{(l)}}(z_{(j)}):={\cal
B}(z_{(j)},\bar{v}_{(l)})=\bigg( \sum_{k=0}^n z_{(j)}^k \bar{v}_{(l)}^k
\bigg)^{1 \over \hbar}. }
Now we have $(f,\Phi_{\bar{v}_{(l)}})=f_{(l)}(v_{(l)}),$ for all $f \in
{\cal F}_{\hbar}$.
The following relation holds 

$$
\Phi_{\bar{v}_{(l)}}(z_{(j)}) = {\cal B}(z_{(j)},\bar{v}_{(l)}) = \exp
\bigg\{ {1 \over \hbar}
{\cal K}(z_{(j)},\bar{v}_{(l)}) \bigg\}, 
$$
\eqn\sesentaiseis{
{\cal K}(z_{(j)},\bar{v}_{(l)})=ln \bigg\{ \sum_{k=0}^n
z_{(j)}^k \bar{v}_{(l)}^k \bigg\}.
}
This relation is known as Berezin's hypothesis A \refs{\berezintwo}.

We intend now to define the covariant symbols of operators acting on
${\cal F}_\hbar$.

Let $\widehat{F}:{\cal F}_\hbar \to {\cal F}_\hbar$ be a linear operator on
${\cal F}_\hbar$. (As for $n< \infty$ the dimension of ${\cal F}_\hbar$ is
finite, every linear operator is also bounded).
Consider the functions $F_B(z_{(j)},\bar{v}_{(l)})$ 
\eqn\sesentaisiete{F_B(z_{(j)},\bar{v}_{(l)}):={(\widehat{
F}\Phi_{\bar{v}_{(l)}},\Phi_{\bar{z}_{(j)}}) \over (\Phi_{\bar{v}_{(l)}}, 
\Phi_{z_{(j)}})},}
for all $j,l.$
These functions are holomorphic on dense subsets $S_{j\bar{l}} \subset
U_j \times \bar{U_l}$ which consist of all points $U_j \times \bar{U_l}$
such that $(\Phi_{\bar{v}_{(l)}},\Phi_{z_{(j)}}) \not= 0$. Moreover, for
any $(p,\bar{q}) \in S_{j\bar{l}} \cap S_{k \bar{m}}$ we have
$$  
F_B(z_{(j)},\bar{v}_{(l)})=F_B(z_{(k)},\bar{v}_{(m)}),
$$
where $(z_{(j)},\bar{v}_{(l)})$ and $(z_{(k)},\bar{v}_{(m)})$ are the
respective coordinates of $(p,\bar{q})$. It means that the set of
functions given by \sesentaisiete\  defines a holomorphic function
$F_B:\cup_{j,l}S_{j\bar{l}} \to \IC$. Observe that
$\cup_{j,l}S_{j\bar{l}}$ is a dense subset of ${\IC}P^n \times
\bar{{\IC}P^n}$. The restriction of the function $F_B$ to the points
$\bar{q}=\bar{p}$ gives an analytic function with respect to the real
structure on ${\IC}P^n$ and is called the {\it covariant symbol} of the
operator $\widehat{F}$. Locally we have   

\eqn\sesentaiocho{F_B(z_{(j)},\bar{z}_{(j)}):={(\widehat{
F}\Phi_{\bar{z}_{(j)}},\Phi_{\bar{z}_{(j)}}) \over (\Phi_{\bar{z}_{(j)}},
\Phi_{z_{(j)}})},} 
for all $j.$
 
Let $f \in {\cal F}_{\hbar}$ be represented in $U_0$ by $f(z)$ and let
$\widehat{F}$ be a linear operator in ${\cal F}_{\hbar}$ then we have

$$
(\widehat{F}f)(z)=(\widehat{F}f,\Phi_{\bar{z}})=(f,\widehat{F}^\dagger 
\Phi_{\bar{z}})
$$

$$
=c(\hbar)
\int_{U_0}(f,\Phi_{\bar{v}})(\Phi_{\bar{v}},\widehat{F}^{\dagger}\Phi_{\bar{z}})\exp\bigg\{-{1 
\over \hbar} {\cal K}(v,\bar{v}) \bigg\} d\mu(v,\bar{v}) 
$$
$$
=c(\hbar)
\int_{U_0}(\widehat{F}\Phi_{\bar{v}},\Phi_{\bar{z}})f(v) \exp \bigg\{-{1 \over \hbar}
{\cal K}(v,\bar{v}) \bigg\} d\mu(v,\bar{v})
$$
$$
=c(\hbar)
\int_{U_0}F_B(z,\bar{v})f(v) \Phi_{\bar{v}}(z) \exp \bigg\{-{1 \over \hbar}
{\cal K}(v,\bar{v}) \bigg\} d\mu(v,\bar{v})
$$

\eqn\sesentainueve{
=c(\hbar)
\int_{U_0} {F_B(z,\bar{v})f(v)(1+z\bar{v})^{1 \over \hbar} \over
(1+v \bar{v})^{{1 \over \hbar} +n+1}} \prod_{k=1}^n {dv^k \wedge
d\bar{v}^k \over 2\pi i \hbar}.
}

Straightforward calculations lead to the following formula for the trace 
of an operator $\widehat{F}$    

\eqn\sesentainueveb{
Tr \  \widehat{F} =c(\hbar) \int_{U_0} F_B(z,\bar{z}) d
\mu(z,\bar{z})=:Tr F_B.}

From the definition of the covariant symbol it follows immediately that
for the unit operator $\widehat{F} =\widehat{1}$, its covariant symbol is
the
unit
function. Using this fact in Eq.  \sesentainueveb\ one finds that the
dimension of the Hilbert space ${\cal F}_{\hbar}$ reads 

$$
dim \ {\cal F}_{\hbar}= Tr \widehat{1} =
c(\hbar)\int_{U_0}
d\mu(z,\bar{z})
$$
\eqn\setenta{
={\Gamma({1 \over \hbar}+n+1) \over \Gamma({1 \over
\hbar}+1) \Gamma(n+1)} =
{{1 \over \hbar} + n \choose {1 \over \hbar}}. }

Finally, if $F_B(z,\bar{z})$ and $G_B(z,\bar{z})$ represent on $U_0$ the
covariant symbols of the operators $\widehat{F}$ and $\widehat{G}$
respectively, than the covariant symbol of $\widehat{F}\widehat{G}$ is
given by the {\it Berezin-Wick star product} $F_B *_B G_B$ which on $U_0$
is represented by

$$
(F_B*_BG_B)(z,{\bar z})= c(\hbar) \int_{U_0} F_B(z,{\bar v})G_B(v,{\bar
z}) {{\cal B}(z,{\bar v}){\cal B}(v,{\bar z}) \over {\cal B}(z,{\bar z})}
\exp \bigg\{-{1 \over \hbar} {\cal K}(v,{\bar v}) \bigg\} d\mu(v,{\bar v})
$$
\eqn\setentaiuno{
= c(\hbar) \int_{U_0} F_B(z,{\bar v})G_B(v,{\bar
z}) 
\exp \bigg\{{1 \over \hbar} {\cal K}(z,{\bar z};v,\bar{v}) \bigg\} 
d\mu(v,{\bar v}), }
where $ {\cal K}(z,{\bar z};v,\bar{v}) :=  {\cal K}(z,\bar{v}) + 
{\cal K}(v,{\bar z}) - {\cal K}(z,{\bar z}) - {\cal
K}(v,\bar{v})$ is the Calabi diastatic function.
As ${\cal K}(z,{\bar z};v,\bar{v})={\cal K}(v,{\bar v};z,\bar{z})$ from
(4.17) and (4.19) it follows that
\eqn\setentaidos{Tr(F_B*_BG_B)=Tr(G_B*_BF_B).}
Hence, the Berezin-Wick star product is a {\it closed star product}
\refs{\connes,\fedosov}.

In order to perform a quantization of the geometric quantum mechanics we
must work with ${\IC}P^{\infty}$ \refs{\koba,\berezinfive}. This can be
done by taking the limit $n \to
\infty$, but one should be careful because some objects might have not
sense at all. For example   
$$
\lim_{n \to \infty} c(\hbar) = \infty. 
$$
The first important result in the case of ${\IC}P^{\infty}$ is that we
still have ${1 \over \hbar}= N \in {\IZ}_+$. Then from (4.18) with $n \to 
\infty$ one gets 
\eqn\sesentaitres{dim \ {\cal F}_{\hbar} = \infty.}
The orthonormal basis of ${\cal F}_{\hbar}$ for $n \to \infty$ can be
chosen analogously as before and it is represented in $U_0$ by the set of
monomials
$$
\bigg\{ e_{(s_1,s_2,...)}(z)=\sqrt{{1 \over \hbar}! \over
({1 \over \hbar}-\sum_{k \not= 0}s_k)!} \prod_{k \not= 0} {(z^k)^{s_k} \over
\sqrt{s_k!}} \bigg\}_{\sum_{k \not= 0} s_k \leq {1 \over \hbar}}. 
$$
As we know from the previous section the general observable on
${\IC}P^{\infty}$ has the form given by (3.2). It seems to be natural to
identify this observable with the Wick symbol of the respective
operator $\widehat{F}$ acting on ${\cal F}_{\hbar}$. So employing (4.16)
and the relation 
\eqn\sesentaitresb{\langle \widehat{\widehat{F}}
\rangle(z,\bar{z})=F_{Wick}(z,\bar{z})=F_B(\bar{z},z)} 
we have in $U_0$

$$
(\widehat{F}f)(z)=c(\hbar)\int_{U_0}{\sum_{k,l=1}^nF_{kl}
z^k \bar{v}^l + \sum_{k=1}^n(F_{0k}\bar{v}^k + F_{k0}z^k)+F_{00}
\over 1 + z\bar{v}} f(v)(1+z\bar{v})^{1 \over
\hbar}(1+v\bar{v})^{-{1 \over \hbar}} d\mu(v,\bar{v})
$$
$$
=\lim_{n \to \infty} \bigg[ (\sum_{k=1}^n F_{k0}z^k + F_{00})
\hbar \lim_{x \to 1}  {\partial \over \partial x} \bigg( x^{1 \over \hbar}
c(\hbar)\int_{U_0}f(v) (1 + {z \bar{v} \over x})^{1 \over \hbar} (1+v
\bar{v})^{-{1\over \hbar}} d \mu (v,\bar{v})
\bigg)
$$
$$
+ \hbar \sum_{l=1}^n \bigg(\sum_{k=1}^nF_{kl} z^k +
F_{0l}\bigg) {\partial f \over \partial z^l} \bigg]
$$

\eqn\setentaicuatro{
=\bigg \{ (F_{00} + \sum_{k=1}^{\infty} F_{k0} z^k) + \hbar
\sum_{l=1}^{\infty} \bigg( F_{0l} - F_{00} z^l + \sum_{k=1}^{\infty}
(F_{kl} - F_{k0} z^l) z^k \bigg) {\partial \over \partial z^l} \bigg \}
f(z),}
where $z \bar{v} := \sum_{k=1}^{\infty} z^k \bar{v}^k$ and we also have used 
the formula $(f, \Phi_{\bar{z}}) = f(z)$.

In particular, substituting (3.3) into (4.23) one gets the Hamilton
operator in the following form

$$
(\widehat{H} f)(z) = \bigg[ \hbar \sum_{k=1}^{\infty}(\omega_k - \omega_0)
z^k {\partial \over \partial z^k} + \omega_0 \bigg] f(z) 
$$
\eqn\setentaicinco{
=\bigg( \hbar \sum_{k=0}^{\infty} \omega_k
\widehat{a}_k^{\dagger}\widehat{a}_k 
\bigg) f(z)
}
where
$$
\widehat{a}_k^{\dagger}\widehat{a}_k=  z^k
{\partial \over \partial z^k},
$$
\eqn\setentaiseis{
\widehat{a}_0^{\dagger}\widehat{a}_0={1 \over \hbar} -
\sum_{k=1}^\infty z^k {\partial \over \partial z^k}.
}

Simple calculations show that the operators defined by \setentaiseis\ can
be extended to the whole Hilbert space ${\cal F}_{\hbar}$ giving the
particle number operators
$\widehat{N}_k:=\widehat{a}_k^{\dagger}\widehat{a}_k$ for $k=0,1,... \ \ .$
{\it However, it is not possible to define in} ${\cal
F}_{\hbar}$ {\it the annihilation} $\widehat{a}_k$ {\it and creation}
$\widehat{a}_k^{\dagger}$ {\it operators}.
This is so because one can not extend globally the operators of the form
${\partial \over \partial z^k}$ and $z^k$.
Using (4.15) and (4.22) one quickly finds that on $U_0$ (in what follows
we omit the subindex Wick to denote the Wick symbol of an operator!)

$$
N_k(z,\bar{z})={1 \over \hbar}{z^k \bar{z}^k \over 1+ z
\bar{z}}, \ \ \ \ \ \  k \not=0,
$$ 
\eqn\setentaisiete{
N_0(z,\bar{z})={1 \over \hbar}{1 \over 1+ z
\bar{z}}. }
The vectors $e_{(s_1,s_2,...)}$ $\in {\cal F}_{\hbar}$ are
eigenvectors of the operators $\widehat{N}_k$

$$
\widehat{N}_k e_{(s_1,s_2,...)}= s_k
e_{(s_1,s_2,...)}, \ \ \ \ \ \ \ \ k \not= 0 
$$
\eqn\setentaiocho{
\widehat{N}_0 e_{(s_1,s_2,...)}=\bigg({1 \over \hbar} - \sum_{k \not= 0} s_k \bigg) e_{(s_1,s_2,...)}. 
}

From (4.25) it follows that the total particle number operator
$\widehat{N}=\sum_k\widehat{N}_k$ has only one eigenvalue: N = ${1 \over
\hbar}$. So
\eqn\setentainueve{\widehat{N}= {1 \over \hbar} \widehat{1}} 
what means that {\it each state is an eigenstate of} $\widehat{N}$ {\it
and the total number of particles is always} ${1 \over \hbar}$.
Now it is clear why we are not able to define annihilation or
creation operators in ${\cal F}_{\hbar}$. This is because the annihilation
of any particle
implies the creation of another one in such a way that the number of
particles is conserved and is equal to ${1 \over \hbar}$.
Of course the vectors $e_{(s_1,s_2,...)}$ are the eigenvectors of
the Hamiltonian (4.24). Namely 

\eqn\ochenta{\widehat{H}e_{(s_1,s_2,...)}= \hbar \bigg( \sum_{k \not= 0}
s_k \omega_k  + ({1 \over \hbar} - \sum_{k \not= 0} s_k) \omega_0 \bigg)
e_{(s_1,s_2,...)}.} 
It follows that {\it the ground state of the field is the state with all}
${1 \over \hbar}$ {\it particles occupying the lowest one particle state}
$\psi_0$ {\it of the energy} $\varepsilon_0$. So the energy of the ground
state is
\eqn\ochentaiuno{E_0={1 \over \hbar}\varepsilon_0} 
and this corresponds to the Bose-Einstein condensation.
As we have decided to identify the functions on ${\IC}P^{\infty}$ with
the Wick symbols rather with the Berezin ones (see (4.22)) we must use the
$*'_B$-product and not the $*_B$-product. Consequently, if $F(z,\bar{z})$
and $G(z,\bar{z})$ are restrictions to $U_0$ of two functions on 
${\IC}P^{\infty}$ which correspond to the field operators $\widehat{F}$
and
$\widehat{G}$, respectively, then the function (the Wick symbol)
corresponding to the product $\widehat{F} \widehat{G}$ is given on $U_0$
by (compare with (2.48))
$$
F(z,\bar{z})*'_BG(z,\bar{z})=c(\hbar) \int_{U_0}
F(v,\bar{z})G(z,\bar{v})\exp \bigg \{ {1 \over \hbar} {\cal
K}(z,\bar{z};v,\bar{v}) \bigg\} d\mu(v,\bar{v})
$$
\eqn\ochentaidos{=G(z,\bar{z})*_B F(z,\bar{z}).}  

In order to consider the star product (4.31) as a formal one (as it is
done in the usual formal deformation quantization) we must expand the
right hand side of (4.31) in the formal series in powers of $\hbar$. This
procedure has been developed in the paper by  N. Reshetikhin and L.
Takhtajan \refs{\takhtajan}. One can easily observe that their normalized
$*$-product (see Eq. (4.6) of \refs{\takhtajan}) is in the present case
exactly the Berezin-Wick $*_B$-product given by Eq. (4.19) because the
unit element 
$e_{\hbar}(z, \bar{z})$ defined by Eq. (4.2) in \refs{\takhtajan} is equal
to the normalization factor $c(\hbar)$ (see Eq. (4.8)) of the present
paper. Therefore using the results of Ref. \refs{\takhtajan} and also the
formulas
$$
c(\hbar)= \hbar^n {\Gamma({1 \over \hbar}+n+1) \over
\Gamma({1 \over \hbar})}=(1+n\hbar)(1+(n-1)\hbar)...(1+\hbar)
$$
$$
1+ \hbar {n(n+1) \over 2} + O(\hbar^2)
$$
and 
$$
A:={1 \over 2} \sum_{j,i\not=0}g^{\bar{j}i} {\partial^2 \over \partial
\bar{z}^j
\partial z^i} \ln \bigg[det(g_{k\bar{l}})\bigg] = - {n(n+1) \over 2} 
$$
one quickly finds that
  
\eqn\ochentaitres{F(z,\bar{z}) *'_B G(z,\bar{z})=G(z,\bar{z}) *_B
F(z,\bar{z})=GF + \hbar \sum_{j,i\not=0} g^{\bar{j}i}{\partial G \over
\partial
\bar{z}^j} {\partial F \over \partial z^i} + O(h^2).}

We must note that in the case when $n \to \infty$, the formal
expansion (4.32) contains divergent terms. Consequently to avoid this
problem we rather use the strict integral formula for the $*'_B$-product
than the formal one. 

From \ochentaitres\ we immediately find that the {\it Berezin-Wick
bracket} defined by
\eqn\ochentaicuatro{\{F,G\}'_B:={1 \over i\hbar}(F*'_BG - G*'_B F) }
reads 
\eqn\ochentaicinco{ \{F,G\}'_B= \{F,G\} + O(\hbar)}
where $\{F,G\}$ is the Poisson bracket of $F$ and $G$
\eqn\ochentaiseis{ \{F,G\}=\sum_{k,l \not= 0} \omega^{\bar{k}l} \bigg({\partial F 
\over \partial \bar{z}^k} {\partial G \over \partial z^l} - {\partial F 
\over \partial z^l}{\partial G \over \partial \bar{z}^k} \bigg)}

\vskip 2truecm

\newsec{Wigner Functions}

In this section we are going to find Wigner functions $\rho_{(s_1,
s_2,...)}(z,\bar{z})$ corresponding to the states $e_{(s_1,s_2,...)}$. In terms of the
Berezin-Wick $*'_B$-product Eq. (4.27) reads 

$$
\big(N_k *'_B \rho_{(s_1,s_2,...)} \big)(z,\bar{z}) =
s_k \rho_{(s_1,s_2,...)}(z,\bar{z}),
\ \ \ \ \ \ k \not= 0
$$

\eqn\ochentaiseis{
\big(N_0 *'_B \rho_{(s_1,s_2,...)}\big)(z,\bar{z}) =
\big({1 \over \hbar} - \sum_{k \not= 0} s_k \big)
\rho_{(s_1,s_2,...)} (z,\bar{z}). }

Employing (4.26) and (4.31) after some work one finds that the system of
equations \ochentaiseis\ is equivalent to the following system of
differential equations

\eqn\ochentaisiete{
\bar{z}^k{\partial \over \partial \bar{z}^k} \bigg[(1+z \bar{z})^{1 \over
\hbar} \rho_{(s_1,s_2,...)}(z,\bar{z}) \bigg] =
s_k  \bigg[(1+z \bar{z})^{1 \over
\hbar} \rho_{(s_1,s_2,...)}(z,\bar{z}) \bigg], \ \ \ \ \ 
k \not= 0. }

(There is no summation over $k$!)

The unique real solution normalized by $Tr \big(\rho_{(s_1,s_2,...)}\big)(z,\bar{z}) =1$, 
where $Tr$ is defined by Eq. (4.17), reads  

$$
\rho_{(s_1,s_2,...)}(z,\bar{z}) = {{1\over \hbar}! \over
({1 \over \hbar} - \sum_{k \not= 0}s_k)! (1+ z \bar{z})^{1 \over \hbar}} 
\prod_{k \not= 0} {|z^k|^{2s_k} \over s_k!}
$$

\eqn\ochentaiocho{
= {e_{(s_1,s_2,...)} (z,\bar{z}) \bar{e_{(s_1,s_2,...)}(z,\bar{z})} \over (1 + z
\bar{z})^{1 \over \hbar}}
}
Hence, the Wigner function $\rho_0$ of the ground state is of the form

\eqn\ochentainueve{
\rho_0(z,\bar{z}) = { 1 \over  (1 + z \bar{z})^{1\over \hbar}}.
}

Then it is easy to find that the expected value $Tr (\widehat{F}
\widehat{\rho}_{(s_1,s_2,...)})$ of any operator
$\widehat{F}$ in the Hilbert space ${\cal F}_{\hbar}$ in terms of the
corresponding Wigner function is given by

\eqn\noventa{
\langle \widehat{F} \rangle = c(\hbar)^2 \int_{U_0} {\rho}_{(s_1,s_2,...)} 
(z,\bar{v}) F(v,\bar{z}) \exp \bigg\{ {1\over \hbar}
{\cal K}(z,\bar{z};v,\bar{v}) \bigg\} d \mu (z,\bar{z}) d \mu(v,\bar{v}).
}

Finally, the von Neumann-Liouville evolution equation for a Wigner
function $\rho(t;z,\bar{z})$ is given by

\eqn\noventaiuno{
{\partial \rho \over \partial t} = \{ \langle \widehat{\widehat{H}}
\rangle, \rho \}'_B.
}

\vskip 2truecm

\newsec{Final Remarks}

In this paper we have investigated the second quantization of the
Schr\"odinger field within the deformation quantization formalism.
Comparing the considerations of Section 2 with the ones of Sections 4 and
5 we
conclude that the Berezin deformation quantization of the geometric
quantum mechanics leads to some results which do not appear at all in the
case of the Berezin deformation quantization of the Schr\"odinger field
(i.e. the usual second quantization). For instance in the former case one
gets that:

(i) $1 \over \hbar$ is a positive integer.

(ii) The number of particles is constant and is equal to $1 \over \hbar$.
Hence, the ground state corresponds to the Bose-Einstein condensation.

(iii) There do not exist the annihilation and creation operators in the
Hilbert space ${\cal F}_{\hbar}$ of the quantized system.

It means that the second quantization and the quantization of geometric
quantum mechanics are not equivalent one to another.

An interesting question is also what happens if we quantize geometric
quantum mechanics corresponding to the nonlinear quantum mechanics ala
Weinberg \refs{\weinberg}. Although difficulties with nonlinear quantum
mechanics seem to be unavoidable (see e.g. \refs{\mielniktwo}), from the
geometric point of view such a quantum mechanics is quite natural
\refs{\anandantwo - \brody}. We are going to study this problem in a
separate paper.

\vskip 2truecm

\centerline{\bf Acknowledgements}
This work was partially supported by the CONACyT (M\'exico) grants 32427E and
33951E and by the KBN (Poland) grant Z/370/S.

\listrefs

\end